\documentstyle[11pt]{article}
\normalbaselines
\begin{document}
\bibliographystyle{unsrt}

\vbox{\vspace{38mm}}
\begin{center}
{\LARGE \bf Selection Rule for Generation Numbers and\\[2mm]
Gauge Anomalies for Unification Groups}\\[5mm]

Huazhong Zhang\\[5mm]
{\it Theoretical Physics Group\\
     Lawrence Berkeley Laboratory\\
     MS 50A-3115, 1 Cyclotron Road\\
     Berkeley, California 94720, USA.}\footnote
{Where the main work supported by DOE} \\[3mm]
{\it and\\
P. O. Box 17660\\
Jackson State University\\
Jackson, MS 39217, USA}\footnote{Permanent address}\\
     (e-mail:fnalv::cpzhang;hzhang@riscman.jsums.edu)\\[5mm]
     (Submitted 25 October 1994)\\[5mm] \end{center}

\begin{abstract}
%insert abstract here
The possibilities of global (non-perturbative) gauge
anomalies for a class of gauge groups are investigated.
Intimately connected to branching rules and topological
aspect of gauge groups, the results are applied to the study
of unification gauge groups such as SO(10), SU(5), $E_6$,
$E_8$ etc. Especially, we discuss extensively about the
selection rule for generation numbers $N_f+N_{mf}=even\ge 4$
in SO(10) and supersymmetric SO(10) unification theories as
originally proposed by the author$^1$, where $N_f$ and
$N_{mf}$ denote the generation numbers for ordinary fermions
and mirror fermions respectively. This is due to the global gauge
anomalies from some subgroups of SO(10) in a fundamental spinor representation
such that the ill-defined 'large' gauge transformations in the subgroup
cannot be unwrapped in SO(10) in the quantum theory as we noted$^1$.
A similar result related to left-right symmetric models is also given.
\end{abstract}

\section{Introduction}
Symmetries and the properties of Lie groups have been always
intriguing in physics, especially in the understanding of
fundamental interactions. In this connection, as a matter of
fact, non-abelian gauge theories have been the main
frame-work in elementary particle physics since the
formulation of Yang-Mills theories$^2$. Lie algebras and
topological properties of Lie groups have played important
roles in non-abelian gauge theories. In particular, the
topological properties of Lie groups have been shedding new
lights on the study of non-abelian gauge theories in the
topological and non-perturbative aspects. In this paper, we
will study some non-perturbative aspects for unification
gauge groups.

Since the building of standard electroweak gauge theory$^3$,
one of the most interesting ideas in particle physics has
been incorporating the standard model into a grand unified
theory$^{4-5}$ (GUT) or a supersymmetric grand unified
theory$^{6-7}$. Although the minimal SU(5) model$^4$
does not lead to the desired unification or is not
compatible with proton decay search$^8$ and CERN LEP
data$^9$, it is now known that the unification may be
achieved in either a supersymmetric GUT, or a GUT with a
gauge group larger than SU(5) spontaneously broken to the
standard gauge group in at least two stages. Such an
example$^{10}$ is the SO(10) GUT model which
may break$^{11}$ first to left-right symmetric
$SU(3)\otimes SU(2)_L\otimes SU(2)_R\otimes U_{B-L}$ model
at a scale $M_X$ and then to the standard model.
Gauge hierarchy problem can be naturally solved in
supersymmetric SO(10) theory, without evidence of
supersymmetry at low energies, non-supersymmetric SO(10)
models are still of interest$^{11}$. The SO(10) models have
many attractive physics features$^{12}$ such that they
preserve a good prediction for the $\sin^2{\theta}_w$ for
Winberg angle, permit neutrinos having small masses through
see-saw mechanism, and may give predictions for some
parameters in the standard model.
One of the interesting features in SO(10) unification models
and supersymmetric SO(10) unification models lies at the
structure of the group representation itself. That is,
the chiral states of quarks and leptons in a family
including a right-handed neutrino can be fitted neatly into
a fundamental spinor representation (f.s) of dimension 16
for the SO(10) gauge group. One of the main purposes of the
present paper is to discuss extensively about this group
theoretic structure and the relevant physics effects.
Especially, we will focus on its global (non-perturbative)
effects. Then, related remarks and consequences will be
given. As a matter of fact, as we will see that the SO(10)
models with Weyl fermions in a fundamental spinor
representation will generate a new type of
global (non-perturbative) gauge anomalies$^1$ when
restricting to some gauge subgroups with non-trivial forth
homotopy group. The ill-defined large gauge transformations
in the subgroups cannot be unwrapped in SO(10).
Therefore, the corresponding generating functional in the
restricted subgroup sector is ill-defined, or the
quantum theory is inconsistent. This also suggests that the
possibilities for this type of global gauge anomalies need
to be taken into consideration
carefully in general in non-abelian gauge theories with Weyl
fermions.

Our discussions will be organized as follows. In the next
section, we will first give a brief description of global
(non-perturbative) gauge anomalies in gauge theories.
Then in section 3, we will clarify the global gauge
anomalies for some gauge groups as the direct product of
SU(2) or SP(2N). In section 4, the results will then be
applied to the discussions of SO(10) as well as other
unification groups and the selection rule for generation
numbers as originally proposed$^1$ by the author. Our
conclusion will be summarized in section 5.

\section{Global (Non-perturbative) Gauge Anomalies in Even
Dimensions}
In this section, we will give a brief description of global
gauge anomalies in even dimensions needed for our
discussions in the paper. We refer the details to the
relevant references.

Gauge anomalies arise as the gauge transformations which are
classically well-defined become inconsistent in the quantum
theory. For a gauge theory with Weyl fermions in 2n
dimensions, if the homotopy group $\Pi_{2n}(G)$ is
non-trivial for the relevant gauge group G, then there can
be topologically non-trivial and continuous gauge
transformations on the compactified spacetime manifold. If
such topologically non-trivial gauge transformations
generate inconsistency in the quantum theory, then the
anomalies are global (non-perturbative). The gauge anomalies
corresponding to the topologically trivial gauge
transformations are perturbative or local.
In general, for a non-abelian gauge theory in an even
dimensions D=2n, one still needs to consider the possibility
of gauge anomalies if the homotopy group $\Pi_{2n}(G)$ for
the gauge group G is non-trivial when the theory is free of
local (perturbative) gauge anomalies.

It was shown by Witten$^{13}$ that an SU(2) gauge theory in
four dimensions with an old number of Weyl fermion doublets
is mathematically inconsistent due to a global
(non-perturbative) gauge anomaly as the sign change of the
fermion measure for the large gauge transformations. With a
global gauge anomaly, the generating functional for the
quantum theory is ill-defined. Topologically, this is
associated with the fact that the homotopy group
$\Pi_4(SU(2))=Z_2$ is non-trivial.
Global gauge anomalies have been investigated
for SU(N) gauge groups$^{14-21}$, and systematically and
rather generally for arbitrary compact and connected simple
gauge groups in generic even dimensions$^{15-21}$,
especially$^{15-21}$ in terms of the James numbers of
Stiefel manifolds and generalized Dynkin indices.
The study of global gauge anomalies are meaningful only if
the gauge theory is free of local (perturbative) gauge
anomalies, since otherwise the theory is anomalous even
for infinitesimal gauge transformations. Furthermore, only
if the infinitesimal gauge transformations are well-defined,
topologically the homotopy group for gauge transformations
can be well-defined since the homotopic equivalence is
defined modulo infinitesimal gauge transformations.

For local gauge anomalies, there can be Green-Schwarz
mechanism$^{22}$ of anomaly cancelation when gravitational
field etc are also included in the theory.
Witten$^{23}$ and others$^{24}$ derived a general formula
for global gauge anomalies including gravitation by index
theorem. But the quantities in the formula in general
do not have a convenient scheme for explicit calculations.
However, for pure gauge theories, one is usually interested
in the strong anomaly cancellation condition $TrX^{n+1}=0$
for a generic Lie algebraic element X for the group under
consideration. The method developed in Refs.14-21 are more
convenient and can be implemented explicitly in many and
rather general cases to calculate the global gauge anomaly
coefficient. In particular, the global gauge anomalies
expressed in terms of$^{16}$ the James numbers of Stiefel
manifold and generalized Dynkin indices
are demonstrated powerful so that many classes of gauge
theories in 2n dimensions have been studied extensively. The
possible global gauge anomalies for many generic groups in
rather generic dimensions have been determined completely in
this approach$^{15-21}$. The essential idea in this approach
can be briefly described as follows. Consider a gauge theory
of gauge group H in 2n dimensions with Weyl fermions in its
representation $\omega$ free of local gauge anomaly. Let
$\Pi_{2n}(H)\neq 0$, so that the theory may possibly possess
a global gauge anomaly. Now consider further a gauge group
G such that H is a subgroup of G and $\Pi_{2n}(G)=\{0\}$.
Then the group G can be regarded as a principal bundle over
G/H with H as the fiber and the structure group.
The fibration $H\rightarrow G\rightarrow $G/H leads to the
exact homotopy sequence$^{25}$
\begin{equation}
...\rightarrow \Pi_{2n+1}(G)\rightarrow \Pi_{2n+1}(G/H)
\rightarrow \Pi_{2n}(H)\rightarrow \Pi_{2n}(G)=0.
\end{equation}
Therefore, the global anomaly of H may be calculated as the
local anomaly of G. It appears as the integration for the
corresponding Wess-Zumino term over a (2n+1)-dimensional
disc $D^{2n+1}$ with the compactified space time $S^{2n}$ as
its boundary. In order for this method to work, one needs
to find a rep. $\tilde{\omega}$ of G such that when G is
restricted to H on the spacetime 2n sphere $S^{2n}$ as the
boundary of the $D^{2n+1}$, the $\tilde {\omega}$ reduces to
the $\omega$ for H plus H singlets. As emphasized in
ref.15-16, usually such a rep condition can be realized
only in the generalized convention of allowing negative
multiplicities for Weyl fermions of different chirality.
For details of the analysis in the approach, see the refs.
14-21. For many groups in arbitrary 2n dimensions, explicit
formulas and rather general results can be obtained in this
approach. As an example, we have obtained the general
formula for the global gauge anomaly coefficient for SU(N)
groups as expressed in the following proposition$^{16}$.\\

{\it Proposition 1}. The global anomaly coefficient
$A(\omega)$ for a rep. $\omega$ of SU(n-k) $(0\le k\le n-2)$
in D=2n dimensions is
given by
\begin{equation}
A(\omega)=exp\{\frac{2\pi i}{d_{n+1,k+1}}Q_{n+1}
(\tilde {\omega})\},
\end{equation}
where
\begin{equation}
d_{n+1,k+1}=\frac{n!}{U(n+1,k+1)}=integers.
\end{equation}
The integral number U(n+1, k+1) is the James number for the
complex Stiefel manifold, SU(n+1)/SU(n-k), and the
$Q_{n+1}(\tilde {\omega})$ is the (n+1)-th Dynkin index for
the $\tilde {\omega}$.

For details of the derivation and discussions of the formula
and James numbers of Stiefel manifolds, see ref.16-20.
For the sake of the later sections, in the following we will
give some of our results$^{15-21}$ as propositions and brief
clarification relevant to our discussions in the present
paper. \\

{\it Proposition 2}. (i) Any irreducible representation
(irrep) $\omega$ of SU(2) has no global anomaly in
D=0(mod 8) dimensions.
(ii) Only spinor reps of SU(2) with spins
$J=\frac{1}{2}(4l+1)=\frac{5}{2},\frac{9}{2},...,$ have
$Z_2$ global anomalies in D=4(mod 8). Neither reps with
$J=\frac{1}{2}(4l+3)=\frac{3}{2},\frac{7}{2},...,$ nor
reps with integer J have global anomalies.

Generally for SP(2N) (N=rank) groups, we have the following
result.\\

{\it Proposition 3}. (i) Any rep $\omega$ of SP(2N) has no
global as well as local gauge anomalies in dimensions
D=0(mod 8)
(ii) Any locally anomaly-free rep $\omega$ of SP(2N) has
no global anomaly in D=2 or 6 (mod 8).
(iii) The global anomaly coefficient A($\omega$) of SP(2N)
in dimension D=4(mod 8) is given by
\begin{equation}
A(\omega ) = exp[i\pi Q_2(\omega )] ,
\end{equation}
where $Q_2(\omega )$ is the 2nd Dynkin index$^{26,27}$
normalized to $Q_2(\Box)=1$ for the 2N-dimensional fundamental rep, and given
by$^{27}$ in general
\begin{equation}
Q_2(\omega)=\frac{d(\omega )}{2N(2N+1)}\sum_{j=1, N}
\{(l_j)^2-(l_j^{(0)})^2\},
\end{equation}
with
\begin{equation}
l_j= f_j+N-j+1, ~l_j^{(0)}=N-j+1, (1\le j\le N),
\end{equation}
for the Young tableau $\Gamma =\{f_1,f_2, ...,f_N\}$
satisfying $f_1\ge f_2\ge ...\ge f_N\ge 0$ corresponding to
$\omega$.

In particular, for SU(2)=SP(2) with N=1, $f_1=2J$
(J=0, 1/2, 1, 3/2...) we have
\begin{equation}
Q_2(\omega )=\frac{2}{3}J(J+1)(2J+1),
\end{equation}

One can easily see that the Proposition 2 is a special case
of the Proposition 3, we list the SU(2) case separately due
to its special interest and importance. \\

{\it Proposition 4}. In arbitrary D=2n dimensions, if the
relevant Weyl fermion rep $\tilde {\omega}$ of G free of
local gauge anomaly in the strong anomaly cancellation
condition $TrX^{n+1}=0$ reduces to an ${\it irreducible}$
rep $\omega$ of H plus H singlets, then there will be no H
global gauge anomalies for the rep $\omega$.

{\it Remark}. Note that the Proposition 4 applies to
semisimple gauge groups H and G also if both the topological
and rep conditions are satisfied at least in the case that
$\Pi_{2n+1}(G$/$H)$ does not contain more than one infinite
cyclic Z's, since there may be only a unique Wess-Zumino
term in this case. The idea of global gauge anomaly
appearing as local in a larger group may determine
effectively the possible global gauge anomalies for the H
gauge group. In the case of more than one Wess-Zumino terms
involved, for instance, if the $\omega$ and
$\tilde{\omega}$ are not an irrep plus singlets, this method
may not apply in general for non-simple groups. An
exceptional case will be seen later in which there are more
than one Wess-Zumino terms involved but the exact homotopy
sequence may be regarded as the direct sum of more than one
due to the fact that the simple ideals in H are effectively
embedded into the corresponding simple ideals of G
respectively, so that each of them for the direct sum of the
exact homotopy sequence then may be regarded as
corresponding to only one topologically independent
Wess-Zumino term for a simple ideal pair in H and G of the
embedding. For a simple gauge group$^{16}$, the above
proposition applies to a generic rep $\omega$ as we have
noted. In the case of $\omega =\sum_i \oplus{\omega}_i$ for a
simple H, the Wess-Zumino term is topologically unique for
the embedding of H into G, although formally it may have
more than one terms from different irreps in the direct sum.

Before going to the next section, we also note the following
facts. It is well known that$^{28, 29}$ local gauge
anomalies can only arise from Weyl fermions in the complex
reps of SU(N) ($N\ge 3$). Topologically, this is due to the
fact that $\Pi_5(G)$ can be infinite cyclic only for G=SU(N)
($N\ge 3$). The groups SO(4k+2) (k=integer) and $E_6$ cannot
have perturbative gauge anomalies although they also have
complex reps, since the $\Pi_5(G)$ is trivial for these
groups G. Furthermore$^{30}$, it is known that for any simple
gauge group with $\Pi_4(G)=\{0\}$ is free of global gauge
anomalies when restricted to any SU(2) subgroup.

\section{Global Gauge Anomalies for Groups as Direct Product
of SU(2) or SP(2N) in D=4 dimensions}
In this section, we will study the possible global gauge
anomalies for a class of semisimple gauge groups as direct
product of SU(2), or SP(2N) more generally. Our present
discussion will be only for the case of D=4 dimensions.
We will present our results in terms of propositions, then
the proofs and remarks will follow.\\

{\it Proposition 5}. The gauge group
$SU(2)\otimes SU(2)\otimes SU(2)$ in the irrep $\omega
=(\Box ,\Box ,\Box )$ or (2,2,2) in terms of dimensions can
have $Z_2$ global gauge anomaly in D=4 dimensions.

Proof. We have H=$SU(2)\otimes SU(2)\otimes SU(2)$ with
the relevant homotopy group
$\Pi_4(H)=Z_2\oplus Z_2\oplus Z_2$. To use our method, we
need to find a group G satisfying the embedding condition,
namely, $H\subset G$ with $\Pi_4(G)=\{0\}$ being trivial,
and G has a irrep $\tilde{\omega}$ such that
$\tilde{\omega}$ reduces to $\omega$ plus H singlets.
It is known that groups SU(N) ($N\ge 8$) contains the
H=$SU(2)\otimes SU(2)\otimes SU(2)$ as a subgroup,
and $\Pi_4(SU(N))=\{0\}$ for those $N\ge 3$.
The branching rule$^{31}$ shows that the fundamental rep
$\tilde{\omega}=(\Box )$ of SU(8) reduces to the
$\omega=(2,2,2)$ upon the reduction of SU(8) to
the $SU(2)\otimes SU(2)\otimes SU(2)$. Therefore, we
can choose G=SU(8). Actually, we can also use
any G=SU(N) ($N\ge 8$). This is due to the fact that$^{16}$
for any rep $\omega$ of SU(N), there is a rep $\omega$' of
SU(N') ($N'> N$) which reduces to the $\omega$ plus SU(N)
singlets. This implies that we can use any G=SU(N)
($N\ge 8$) in its fundamental rep. However, for simplicity,
we will use G=SU(8).
Then the possible $H=SU(2)\otimes SU(2)\otimes SU(2)$ global
gauge anomalies will appear as the Wess-Zumino
term$^{14-16}$ corresponding to the embedding of H into
G=SU(8) in the fundamental representation. The corresponding
exact homotopy sequence for the fibration $H\rightarrow
G\rightarrow G$/$H$ is given by$^{14-20}$
\begin{equation}
...\rightarrow \Pi_5(G)\rightarrow \Pi_5(G/H)\rightarrow
\Pi_4(H)\rightarrow \Pi_4(G)=0.
\end{equation}
With $H=SU(2)\otimes SU(2)\otimes SU(2)$ and G=SU(8), this
is written as$^{25}$
\begin{equation}
...\rightarrow Z\rightarrow Z\oplus Z_2\oplus Z_2
\rightarrow Z_2\oplus Z_2\oplus Z_2\rightarrow 0.
\end{equation}
We can calculate the formula for the basic global anomaly
coefficient is this case, and it is given by
\begin{equation}
A=exp\{i\pi bQ_3(\Box )\}
\end{equation}
with b being an integer,  or this can be rewritten as
\begin{equation}
A=exp\{i\pi bQ_2(\Box )\}
\end{equation}
with the even odd rule$^{30,15-20}$
$Q_2(\tilde{\omega})=Q_3(\tilde{\omega})$ (modulo 2)
for any $\tilde{\omega}$ of SU(N).
The $Q_2(\Box )$ and $Q_3(\Box )$ are the second and third-order Dynkin indices
for the fundamental representation $\Box$ of SU(8) respectively with
the possible constraint that the SU(8) gauge theory should be free of
local anomaly when restricting to the three SU(2) factors on
the spacetime $S^4$ as the boundary of a five-dimensional
disc $D^5$. However, since the three SU(2) factors are
automatically free of local gauge anomaly in four
dimensions, there is no constraint on the Dynkin index.
The integer b is odd or even depending on whether the odd
topological number for the Z in $\Pi_5(G/H)$ (up to finite
cyclic part) can be mapped to some non-trivial element in
$\Pi_4(H)$ or not. It is known that the three $Z_2$'s in the
$\Pi_4(H)$ are completely symmetric, the subgroups split in
a canonical way, there is a special element $h_1\oplus
h_2\oplus h_3$ with the $h_i$ (i=1,2,3) being the generators
for the three $Z_2$'s respectively. The odd elements of Z in
the $\Pi_5(G/H)$ are then indeed mapped to the
$h_1\oplus h_2\oplus h_3$. We note that in order to obtain
the odd topological number for Z in the
in $\Pi_5(G/H)$, the corresponding gauge transformation
on the $D^5$ needs to be topologically non-trivial in all
the three SU(2) factors when restricting to its boundary
$S^4$. Only in this case, none of the SU(2) factors may be
topologically reduced or equivalently to give a factor 2
from the dimension as a multiplicity for the other SU(2)
factors, so that the eight-dimensional (irreducible)
embedding is topologically effective. This is also because
the relevant $\Pi_4(H)$ has one more $Z_2$ than the torsion
of $\Pi_5(G$/$H)$. A more rigorous proof needs Steenrod
algebra and Postnikov systems$^{25}$ and is
too involved to be given here. We are only interested in the
result here. Thus, with $Q_2(\Box)=1$, the basic global
anomaly coefficient in this case is then $A=-1$. Therefore,
the $\omega =(2,2,2)$ of $SU(2)\otimes SU(2)\otimes SU(2)$
can have $Z_2$ global gauge anomaly in D=4 dimensions.\\

{\it Remark}. The fact that the $h_1$, $h_2$ and $h_3$ above
do not generate global anomalies does not necessarily imply
that the $h_1\oplus h_2\oplus h_3$ cannot generate a global gauge
anomaly, since the relevant homomorphism induced by the exact
homotopy sequence is from the $\Pi_5(G$/$H)$ to the $\Pi_4(H)$
but not reversely.\\

{\it Proposition 6}. The gauge group $SU(2)\otimes SU(2)$
in the irrep $\omega =(\Box ,\Box )$ or (2,2)
in terms of dimensions can have $Z_2$ global gauge
anomaly in D=4 dimensions.

The proof is essentially the same as that for Propostion 5,
except that now H=$SU(2)\otimes SU(2)$, G=SU(N) ($N\ge 4$)
in its fundamental rep. For simplicity, we can choose
G=SU(4) in the fundamental rep $\tilde{\omega}=(\Box )$ which
reduces to the (2,2) rep of H upon the reduction of
$G\downarrow H$. The relevant exact homotopy sequence is
\begin{equation}
...\rightarrow Z\rightarrow Z\oplus Z_2\rightarrow Z_2
\oplus Z_2\rightarrow 0.
\end{equation}
The $\Pi_4(H)$ again has one more $Z_2$'s than that in
the $\Pi_5(G$/$H)$. The global gauge anomaly coefficient is
this case is $A=exp\{i\pi Q_2(\Box )\}=-1$ with
$Q_2(\Box )=1$ for SU(N) ($N\ge 4$).\\

{\it Remark}. An immediate implication of this proposition is that
a left-right symmetric model with an odd number of (2,2) Weyl fermions
in $SU(2)\otimes SU(2)$ is inconsistent due to a global
gauge anomaly.\\

{\it Proposition 7}. The gauge group
$SU(2)\otimes SU(2),...,\otimes SU(2)$ as N SU(2) factors
with $N\ge 4$ in the irrep
$\omega =(\Box ,\Box ,...,\Box )$
or (2,2,...2) in terms of dimensions have no global gauge
anomaly in D=4 dimensions.

Proof: This can be seen in two different cases depending on
N=even or odd. If N=2k=even ($\ge 4$), then we can regard
the H as k $(k\ge 2)$ factors of $SU(2)\otimes SU(2)$ in
((2,2),(2,2),..,(2,2)) rep. Since each $SU(2)\otimes SU(2)$
factor can be embedded into SU(4) in its fundamental rep
$\Box$ of dimension 4 which reduces to (2,2) upon reduction
$SU(4)\downarrow SU(2)\otimes SU(2)$. We can choose
G=$SU(4)\otimes SU(4)...\otimes SU(4)$ as k factors of SU(4)
in the irrep (4,4,....,4). In this case, the homotopy group
$\Pi_5(G$/$H)=\sum_{1}^{k} \oplus Z\oplus Z_2$. The
embedding of H into G leads to k independent Wess-Zumino
terms corresponding to the fact that each $SU(2)\otimes
SU(2)$ is embedded into a corresponding SU(4). Therefore, the
corresponding exact homotopy sequence is decomposed into
the direct sum of k independent ones. Each of them has the
same structure as that in the Proposition 6. However, the
Lie algebra representation for each of the SU(4) factors now
is obviously no longer irreducible, since an SU(4) Lie
algebra is now from the restriction of that for the k SU(4)
factors to a single one. The G as k SU(4) factors is in the
rep (4,4,...,4), the Lie algebra elements, say for the first
SU(4) factor are of the form $\{L_a\otimes (1)^2_{4,4}
\otimes ...\otimes (1)^k_{4,4}\}$ with $(1)^i_{4,4}$ being
the $4\times 4$ unit matrix corresponding to the ith SU(4)
subalgebra, where the $\{L_a\}$ are the Lie algebra of SU(4)
in the fundamental irrep. Let us recall that$^{14-16}$ the
Wess-Zumino term is enclosed by a trace operation for the
matrix rep of the Lie algebra of G, this guarantees that
the global anomaly coefficient is in the form of
$A=exp(i\pi 2m)=1$ (m=integers). Therefore, the theory has
no global gauge anomalies for $N=2k\ge 4$. One can also
formally see this
by regarding the embedding as corresponding to only one
Wess-Zumino, then the trace operation in G automatically
splits into a summation of k independent terms, each of
them corresponds to an independent Wess-Zumino term for the
embedding for $SU(2)\otimes SU(2)$ into a SU(4), except that
the rep is no longer irreducible as we have seen above.
When $N=2k+1\ge 5$, then one can choose G as the semisimple
gauge group consisting of k SU(4) factors and one SU(8)
factors in the irrep $(\Box ,...,\Box )$ or (4,4,...,4,8) in
terms of dimensions. Note that each dimension factor
corresponding to the simple ideal is even, for the same
reason as above, the global anomaly coefficient must be 1,
the theory is free of global gauge anomaly.\\

{\it Proposition 8}. The gauge group
$SP(4)\otimes SP(4)$ in the irrep $\omega =(\Box ,\Box )$
or (4,4) in terms of dimensions have no global gauge anomaly
in D=4 dimensions.

Proof: This can be seen easily by using quite different
choices of gauge group G. Let us first see this by using
G=SO(10) in a fundamental spinor rep. of dimensions 16.
The branching rule$^{31}$ shows that a rep 16 of SO(10)
reduces to (4,4) upon the reduction of
$SO(10)\downarrow SP(4)\otimes SP(4)$, and we also have
$\Pi_4(SO(10))=\{0\}$. Then according to our Proposition
4, the Proposition 8 is immediate. We can also see this
by using $G=SU(4)\otimes SU(4)$ in the irrep (4,4) with
$\Pi_4(SU(4)\otimes SU(4))=\{0\}$.
The branching rule$^{16, 31}$ shows that it reduces to the
(4,4) for $SP(4)\otimes SP(4)$ upon the reduction of G to
it. Then the same proof as that in Proposition 7 applies.\\

{\it Proposition 9}. The gauge group
$SP(2N_1)\otimes SP(2N_2)\otimes ... \otimes SP(2N_k)$
with $N_i\ge 2,~k\ge 2$ in the irrep $\omega=(\Box ,\Box
,...,\Box )$ or $(2N_1,2N_2,...,2N_k)$ in terms of dimensions
have no global gauge anomaly in D=4 dimensions.

Proof. We first note the fact that $SP(2N)\subset SU(2N)$,
and the branching rule$^{31}$ that (2N) reduces to (2N) upon
the reduction $SU(2N)\downarrow SP(2N)$. Then the
proposition can be easily shown by using
G=$SU(2N_1)\otimes SU(2N_2)\otimes ... \otimes SU(2N_k)$
in the irrep $\omega=(\Box ,\Box ,...,\Box )$
or $(2N_1,2N_2,...,2N_k)$ in terms of dimensions with
$\Pi_4(G)=\{0\}$. Same as in the proof of the Proposition 7,
the theory is free of global gauge anomaly.

We have shown several propositions for the possibilities of
global gauge anomalies of a semisimple gauge group with
more than one SP(2N) ideals. As we will see that some of
the above results will be useful to the study of unification
gauge groups.

\section{Gauge Anomaly for SO(10) and Other Unification
Groups and the Selection Rule for Generation Numbers}
     The global (non-perturbative) gauge anomalies relevant
here for SO(10) unification group are more subtle than the
usual ones similar to that first noted by Witten$^{13}$ with
the non-trivial homotopy group $\Pi_4(G)$ for the
gauge group G in four dimensions. Since $\Pi_4(SO(10))={0}$
is trivial, the usual expectation is that there should not
be global gauge anomalies. Our idea is to consider
the subgroups of SO(10) with non-trivial forth homotopy
group. Classically, such non-trivial topological structures
are unwrapped in SO(10). In quantum
theory, however, if the non-trivial topological structure
can generate gauge anomalies, the corresponding large gauge
transformations in the subgroup are
anomalous or ill-defined. Obviously, the unwrapping then
can not be physically well-defined since an ill-defined
symmetry transformation in quantum theory
cannot be homotopically equivalent to the identity
transformation which is always well-defined.

The new global gauge anomalies we noted$^1$ arise from the
restriction of a gauge group G with relevant trivial
homotopy group to its gauge subgroups (containing more than
one simple ideals with non-trivial forth homotopy group).
The example we will discuss is the subgroups of SO(10) due
to its crucial importance and relevance to the unification
theories. We will now describe our result for the SO(10)
gauge theories, and may use the Lie algebras for the
discussion of representations. The same notations may be
used for the Lie groups and corresponding
Lie algebras, no confusion should be caused in our
discussion here. We will first show the following
proposition.\\

{\it Proposition 10}. The SO(10) gauge group with Weyl
fermions in a sixteen-dimensional fundamental spinor (f.s)
rep can have a $Z_2\oplus Z_2$ global gauge anomaly when
restricted to the
$SU(2)\otimes SU(2)\otimes SU(2)\otimes SU(2)$ subgroup
obtained through the reduction of its subgroup
$SU(2)\otimes SO(7)$ with SO(7) to the subgroup
$SU(2)\otimes SU(2)\otimes SU(2)$.

The SO(10) group contains a maximal subgroup $SU(2)
\otimes SO(7)$ (The difference between SU(2) and SO(3) in
our consideration is immaterial since they have the same
forth homotopy group). Restricting the SO(7) to its
subgroup $SU(2)\otimes SU(2)\otimes SU(2)$, then we obtain a
subgroup $SU(2)\otimes SU(2)\otimes SU(2)\otimes SU(2)$.
For this subgroup, we have the relevant
homotopy group
\begin{equation}
\Pi_{4}(SU(2)\otimes SU(2)\otimes SU(2)\otimes SU(2))
=Z_2\oplus Z_2\oplus Z_2\oplus Z_2.
\end{equation}
The homotopy group topologically classifies the continuous
gauge transformations restricted to this subgroup in the
compactified spacetime manifold.
The non-trivial topological $Z_2\oplus Z_2\oplus Z_2
\oplus Z_2$ structures exist when the SO(10) gauge theory is
restricted to the subgroup. Such a non-trivial
topological structure can be unwrapped classically.
However, as we have stressed that if the gauge
transformations in the subgroup can be anomalous, then
in the quantum theory, the gauge transformations in the
subgroup are ill-defined, and such an unwrapping cannot be
well-defined. We will show that this is indeed the case, and
therefore, the theory has a non-perturbative gauge anomaly.
The meaning of the global (non-perturbative) gauge anomaly
here may be regarded as that the corresponding gauge
transformations cannot be continuously deformed into
identity transformations in quantum theory.

The branching rule$^{31}$ for a fundamental spinor
representation (f.s) or (16) of SO(10) in the above reduction
$SO(10)\downarrow SU(2)\otimes SU(2)\otimes
SU(2)\otimes SU(2)$ can be written as
\begin{equation}
(16)\rightarrow (2-1-2-2)\oplus (2-2-1-2)
\end{equation}
in terms of dimensions.
To determine the possible global anomalies, we will first
consider an irreducible representation (2-1-2-2) in the
above branching, then the overall possibilities can be
clarified. Obviously, for the irreducible representation
(2-1-2-2), it is equivalent to consider the possible global
anomaly for the group as three SU(2) factors in the
irreducible representation (2-2-2), since the Weyl fermions
are invariant under the gauge transformations restricted in
the second SU(2) gauge group. Now according to our
Proposition 5, such a irrep has a $Z_2$ global gauge
anomaly. Thus, the first irrep (2,1,2,2) in the branching
rule can contribute a $Z_2$ anomaly if it is not canceled
by the second irrep (2,2,1,2). For the same reason, the
(2,2,1,2) can also have a $Z_2$ anomaly. Whether the two
$Z_2$ anomalies cancel or not depends on if they have to
arise from topologically equivalent gauge transformations in
the gauge subgroup $SU(2)\otimes SU(2)\otimes SU(2)\otimes
SU(2)$. It is obvious that the two large gauge
transformations in the subgroup are topologically
inequivalent if the first one is topologically non-trivial
simultaneously in the first, third, and forth SU(2) ideals
in our notation, but the second one is topologically
non-trivial simultaneously in the first, second, and forth
SU(2) ideals. Namely the two gauge transformations of
topological numbers (1,0,1,1) and (1,1,0,1) corresponding
to the homotopy group $Z_2\oplus Z_2\oplus Z_2\oplus Z_2$
generate two independent $Z_2$ anomalies. Therefore, the
SO(10) theory with Weyl fermions in a fundamental spinor rep
when restricted to the $SU(2)\otimes SU(2)\otimes
SU(2)\otimes SU(2)$ obtained through the reduction as in the
above proposition can have $Z_2\oplus Z_2$ global gauge
anomalies. The Proposition 10 is then proved.

The consequence is that$^{13}$ the generating functional and
the operators invariant under such gauge subgroup cannot be
well-defined relative to the relevant large and continuous
gauge transformations in the subgroup.
The $Z_2\oplus Z_2$ anomaly may be understood
as that when the gauge transformation is topologically
non-trivial in three of the SU(2) factors simultaneously but
trivial in either the second or the third
one in our notation, the fermion measure will change a sign,
the quantum theory is then not well-defined$^{13}$.

{\em Remark}. The two different $Z_2$ global gauge anomalies
for the $Z_2\oplus Z_2$ arise from the two different
irreducible representations in the branching rule eq.(12).
They correspond to topologically inequivalent
gauge transformations when restricting to the relevant
subgroup. As we emphasized, since some 'large' gauge
transformations in the subgroup are ill-defined or
anomalous, they cannot be unwrapped to the identity in
SO(10) in the quantum theory. Therefore, the fact that the
SO(10) gauge group with $\Pi_4(SO(10))=\{0\}$ does not have
local gauge anomalies will not contradict
to our result. Note that for each of the
irreducible representations in eq.(12) (e.g the (2,1,2,2)
for the four SU(2) factors), it cannot be
embedded into a representation $\tilde{\omega}$ of SO(10)
such that the $\omega$ reduces to the irreducible
representation (2,1,2,2) plus singlets upon the
reduction. This is also an explicit example showing that the
conventional proposition$^{16}$ noted by using the
Wess-Zumino term argument does not apply generally
to the case in which the relevant subgroup has more than
one ideals with non-trivial 2n-th homotopy group in D=2n
dimensions and the representation is not irreducible.
As we have seen that this is also why a fundamental
representation (f.s) of the SO(10) cannot have $SP(4)
\otimes SP(4)$ global anomaly, due to the fact that the
(f.s) reduces to the sixteen-dimensional {\it irreducible}
representation ($\Box , \Box $) upon the reduction
$SO(10)\downarrow SP(4)\otimes SP(4)$, the conventional
argument of using Wess-Zumino term for the SO(10) may apply.
In this case, the vanishing of SO(10) local anomaly or
Wess-Zumino term implies the absence of the relevant
$SP(4)\otimes SP(4)$ global anomaly.
The problem with embedding a direct sum with more than
one irreducible representations of global gauge
anomalies is that the anomaly information may not be
extracted independently due to the fact that$^{15-16}$ the
global gauge anomaly for an irreducible
representation free of local gauge anomaly can be at most of
$Z_2$ type. Generally, from this point of view for the
global gauge anomalies, the restriction of a gauge theory to
a gauge subgroup H may not be the same as embedding the
gauge subgroup H into the original gauge group G due to the
representation condition needed to extract the possible
global gauge anomalies.

{\it Remark}. We have also noted that the $SU(2)\otimes
SU(2)\otimes SU(2)\otimes SU(2)$ up to isomorphism is the
only possible subgroup of SO(10) in a (f.s) rep having
$Z_2\oplus Z_2$ global gauge anomalies. Another example is
the $SU(2)\otimes SU(2)\otimes SU(2)\otimes SU(2)$
subgroup obtained through the reduction of the subgroup
$SU(2)\otimes SU(2)\otimes SO(6)$ of SO(10) with the SO(6)
to two SU(2) factors.

Note also that generally gauge symmetry in a gauge group
implies the gauge symmetry in its gauge subgroup
(see ref.32 for the other studies related to this
property), namely a well-defined gauge theory needs to be
well-defined when restricting to its gauge subgroups. In
quantum theory, if there are gauge anomalies when
restricting to a gauge subgroup, then the gauge theory
cannot be well-defined. In conclusion, the SO(10)
gauge theories with Weyl fermions in a fundamental spinor
representation of dimension 16 have global
(non-perturbative) gauge anomalies. The SO(10) has two
fundamental spinor representations which are
complex conjugate to each other. Our results applies to
either one of them.

Denote the numbers of two inequivalent fundamental spinor
representations
as $N(16)$ and $N(\bar{16})$, obviously we need to have
\begin{equation}
N(16)+N(\bar{16})=even,
\end{equation}
in order to cancel out the global gauge anomalies.
Consequently, SO(10) unification models with three
generations of fermions have global gauge anomalies.
We have checked that the adjoint representation of
dimensions 45 for the SO(10) will be free of global gauge
anomalies for the relevant subgroups. Therefore, our
conclusion applies both to the non-supersymmetric SO(10)
models and supersymmetric models in which gauginos are in
the adjoint representation.

The physics consequences of our result may be of fundamental
interest if the SO(10) gauge theories are relevant to the
realistic world. Obviously, the SO(10) models and
supersymmetric SO(10) unification models need to be modified
according to our analysis. In the usual physics
convention for the Weyl fermions with the observed three
families of leptons and quarks, we have the following
selection rule.\\

{\it Selection Rule For Generation Numbers}\\
In SO(10) and supersymmetric SO(10) unification models,
the Weyl fermions need to obey the selection rule written as
\begin{equation}
N_f+N_{mf}=even\geq 4,
\end{equation}
with the $N_f=N(16)$ and $N_{mf}=N(\bar{16})$ denoting the
number of fermion families and the number of mirror fermion
families respectively.

Therefore, we predict that there will be at least one more
fermion family or at least one mirror fermion family if an
SO(10) unification gauge theory is realistic. Where in the
content of SO(10) unification, the fourth generation (or a
generation of mirror fermions) also includes
a right-handed neutrino (or a left-handed mirror neutrino).
Mirror fermions have the
same $SU(3)\otimes SU_L(2)\otimes U(1)_Y$ quantum numbers as
the ordinary fermions except that they have opposite
handedness. Usually$^{33}$, mirror fermions are considered
with three generations. Conventionally, one
family of mirror fermions seems not so motivated. However,
our result of the global gauge anomalies shows that it is
one of the simple ways to cancel the global anomalies. As in
the usual discussions, if there exists fourth
generation of fermions with V-A weak interaction, then of
course, it seems natural to have either no mirror fermions
or four families of mirror fermions. The next possibility is
either to have three generations of ordinary fermions and
three generations of mirror fermions correspondingly as in
the usual discussions of mirror fermions,
or to have six generations of fermions with three more
repetitions of an ordinary fermion family. If there are
mirror fermions, one of the
most fundamental consequences will then be that the Lorentz
structure of the weak interaction will no longer be chiral
with only V-A currents coupling to the W gauge bosons, there
will be also V+A piece which though may be very small
relevant to the current experimental observation$^{33}$.
There has been analysis$^{33}$ about the charged and neutral
current data suggesting that the possible V+A impurity in
the weak amplitudes is typically less that about 10\%.

We will now give a brief sketch of some other related
physics issues, for details see the relevant references.
In the content of the electroweak theory, either an
additional generation of fermions or a generation (three
generations) of mirror fermions obtain their masses through
the electroweak symmetry breaking at the order of
about O(300Gev), this will give effects on low energy
physics and also subject to both theoretical and
experimental constraints. The LEP date set a lower
bound for their masses denoted by $M_F$ at about $M_F\geq
m_z/2$, namely about half of the Z boson mass. The partial
wave unitarity$^{34}$ at high
energies shows that the masses above about
$O(600Gev$/$\sqrt{N_{DQ}})$ and $O(1Tev$/$\sqrt{N_{DL}})$
for quarks (or mirror quarks) and leptons (mirror leptons)
will signal the breakdown of
the perturbation theory, where $N_{DQ}$ denotes the
total number of nearly degenerate weak-isospin doublets for
quarks and mirror quarks, and similarly with $N_{DL}$ for
leptons and mirror leptons. There may be stringent
constraint on the masses, mass splitting in a
weak-isospin doublet for possible new fermions due to the
bound on the correction $\delta\rho$ for the
parameter$^{35,36}$
$\rho ={m_w}^2$/${m_z}^2\cos^{2}{\theta}_{w}$
from its tree level value in the minimal standard model,
as well as for the other precision electroweak
parameters$^{35, 37}$. It is known that the radiative
corrections$^{35}$ in perturbation theory can play an active
role in this. There are also recent discussions that$^{36}$
the possible bound states formed by the exchange
of Higgs bosons in the presence of additional heavy fermions
may give non-perturbative contribution $\delta \rho<0$ and
cancel the perturbative correction within the current
experimental error in the nearly degenerate case,
and therefore can relax the constraints for the masses and
mass splitting due to the $\rho$ parameter.

An additional family of fermions or mirror fermions may also
be constrained by that Yukawa couplings should remain small
during the evolution in the perturbative region
(about $\alpha_{Yuk}={\lambda_{Yuk}}^2$/$4\pi\leq 1$),
otherwise its running may induce Landau poles in the
one-loop approximation. The presence of these singularities
at some scale signals the breakdown of perturbation
theory and the probable triviality of the continuum limit.
Related to the running of Yukawa couplings and infrared
fixed-point solution$^{38}$ to the renormalization group
equations, there has been discussions$^{38, 39}$
in supersymmetric unification models with Yukawa coupling
unification (e.g. $\lambda_{\tau}=\lambda_b$ at the
unification scale) that only small regions in the
$m_t-\tan\beta$ ($\tan\beta=v_{up}$/$v_{down}$) plane may
be allowed (e.g. about $1\leq\tan\beta\leq 1.5$ or
$\tan\beta\geq 40\pm 10$ with $m_t\leq 175 Gev$). The value
of $\tan\beta$ can typically effect$^{38-39}$ the flavor
changing neutral currents in processes like
$b\rightarrow s\gamma$ and $B\bar{B}$ mixing and to the
proton decay$^{40}$.
It also constraints on the Higgs boson masses$^{41}$ which
may be relevant to the LEP II. If there are additional
fermions or mirror fermions, one may expect that their
presence will also have these typical effects,
relevant discussions with Yukawa coupling unification at the
unification scale then may need to incorporate them. At
least, the possibility of one additional generation of
mirror fermions from our motivation in terms
of global gauge anomalies sounds quite new.
It has been argued$^{42}$ in the conventional mirror fermion
models (with three generations of mirror fermions
corresponding to ordinary observed generations of quarks and
leptons) that mirror doublets should always be
assumed degenerate in masses (without considering the
non-perturbative effects in ref. 36) in order to reproduce
the precision LEP data, and the possible Higgs masses may be
in rather restricted regions. We note that mirror fermions
of at least three generations usually appear in the particle
spectrum of many theories other than some
superstring models with family unification, such as those
with extended supersymmetry ($N\geq 2$) imposed on a gauge
theory, in the Kaluza-Klein theories$^{43}$, and some
composite models$^{44}$. However, according to our analysis
of global anomalies in SO(10) unification gauge theories,
one of the interesting models is to have only one generation
of mirror fermions besides the three generations of ordinary
fermions. In general, one may expect that$^{33,42}$
mirror fermions need to mix with the ordinary fermions in
order to avoid stable mirror fermions although the mixing
may be small. If there exists only one generation of mirror
fermions, fundamentally, it is unnatural to assume that it
corresponds to a particular family of ordinary fermions.
Therefore, this generation of mirror fermions
will mix with all the three generations of ordinary
fermions, and this then will induce the flavor mixing
between the three generations of ordinary
fermions also, this seems to provide another origin for the
possible flavor mixing and possible CP violations. Moreover,
as it is known that the lifetime of heavy neutrinos may
subject to cosmological constraint$^{45}$
(a suggestion is about $\tau < 10^3 yr (1kev/m_v)^2$) since
it may be strongly believed that the age of the universe is
greater than $10^{10}$ years. Our selection rule may have
consequences on the structure of the universe.
Our rule for the generation numbers can be of fundamental
interest and importance. We will conclude this section with
the following related propositions and discussions. \\

{\it Proposition 11}. The SO(10) gauge group with Weyl
fermions in a fundamental spinor (f.s) rep can have a
$Z_2\oplus Z_2\oplus Z_2\oplus Z_2$ global
(non-perturbative) gauge anomaly when restricting to
$SU(2)\otimes SU(2)\otimes SU(2)\otimes SU(2)\otimes$
obtained through the reduction of the subgroup
$SP(4)\otimes SP(4)$ of SO(10) with each SP(4) to two SU(2)
factors.

Proof: The reduction of SO(10) to its subgroup
$SP(4)\otimes SP(4)$ has the branching rule$^{31}$
$(f.s)\rightarrow (4,4)$. Upon the reduction of SP(2)
to $SU(2)\otimes SU(2)$, we have
$(4)\rightarrow (1,2)\oplus (2,1)$. Thus, the branching
rule for the reduction $SO(10)\downarrow
SU(2)\otimes SU(2)\otimes SU(2)\otimes SU(2)\otimes$
in this case is given by
\begin{equation}
(16)\rightarrow (1,2,1,2)\oplus (1,2,2,1)
\oplus (2,1,1,2) \oplus (2,1,2,1) .
\end{equation}
According to the proposition 6, each of the irrep in the
branching rule above may contribute a $Z_2$ anomaly.
The theory obviously has
$Z_2\oplus Z_2\oplus Z_2\oplus Z_2$ global gauge anomalies
corresponding the gauge transformations with the topological
numbers (0,1,0,1), (0,1,1,0), (1,0,0,1), (1,0,1,0)
respectively. The proposition is then proved.

{\it Remark}. Obviously, the $Z_2\oplus Z_2\oplus Z_2\oplus
Z_2$ global gauge anomalies for the SO(10) in a fundamental
rep cancel out if the total number of families for fermions
and mirror fermions is even. Therefore, although the SO(10)
can have different types of global gauge anomalies, but they
all cancel out if our selection rule for the generation
numbers is satisfied.\\

{\it Proposition 12}. The vector representation of dimension
10 for the SO(10) can have global gauge anomalies when
restricting to certain semisimple subgroups as product of
SU(2) factors through many reductions. But they all cancel
out if the total number of the vector representations are
even. Some examples are given below.\\
(1)$Z_2\oplus Z_2$ global anomalies when restricting to the
$SU(2)\otimes SU(2)\otimes SU(2)\otimes SU(2)$
through the reduction of subgroup $SP(4)\otimes SP(4)$ of
SO(10) with each SP(4) to two SU(2) factors;\\
(2) $Z_2\oplus Z_2$ global gauge anomalies when restricting
to a $SU(2)\otimes SU(2)$ gauge subgroup through the
following reductions:\\
(2a)The $SU(2)\otimes SU(2)$ from the reduction of either
one of the SP(4) factors in (1);\\
(2b) the $SU(2)\otimes SU(2)\otimes SO(6)$
($SO(6)\cong SU(4)$) subgroup of SO(10) to the two SU(2)
factors;\\
(2c) by the reduction of the subgroup $SU(2)\otimes SO(7)$
of SO(10) to $SU(2)\otimes SU(2)\otimes SU(2)\otimes SU(2)$,
the relevant two SU(2) factors both are in the fundamental
representation in one of the irreducible representations
(2-2-1) in the branching rule for the reduction of the SO(7)
to $SU(2)\otimes SU(2)\otimes SU(2)$.
Moreover, the SO(10) group in low-dimensional
representations of dimensions 45, 54, 120 etc. will not have
global gauge anomalies.

By using the branching rules$^{31}$ for these reductions,
and our propositions in section 3, it is straightforward to
verify the above proposition. We only show this for the case
(1) as an example. Upon the reduction
$SO(10)\downarrow SP(4)\otimes SP(4)$, we have
$(10)\rightarrow (5,1)\oplus (1,5)$. Since
$(5)\rightarrow (2,2)\oplus (1,1)$ upon
$SP(4)\downarrow SU(2)\otimes SU(2)$. This gives
$(10)\rightarrow (2,2,1,1)\oplus (1,1,2,2)\oplus 2(1,1,1,1)$
upon the reduction
$SO(10)\downarrow SU(2)\otimes SU(2)\otimes SU(2)
\otimes SU(2)$. According to the proposition 6, the theory
restricted to the $SU(2)\otimes SU(2)\otimes SU(2)
\otimes SU(2)$ obviously can have $Z_2\oplus Z_2$ gauge
anomalies.

We especially emphasize that since in the supersymmetric
SO(10) models, the gauginos are in the 45-dimensional
adjoint representation which is free of gauge anomalies,
our selection rule applies both to
non-supersymmetric SO(10) and supersymmetric SO(10)
unification theories. Before the summary of our conclusions,
we will also briefly discuss about some other gauge groups
relevant to unification theory. It is straight to verify the
results with the relevant branching rules and our
propositions in the section 3.

{\it Remark}. Obviously, the global gauge anomalies in a
fundamental spinor representation cannot be canceled by
adding more vector representations and vice versa.
It is a typical feature that the branching rules become much
more involved when the dimension goes higher, if there were
other possible anomalies they would be very dependent on the
reduction procedure, global anomalies from representations
of different dimensions may not cancel each other.
At least, we checked up to dimensions of several hundred, no
other higher irreducible representations can have the same
possibilities for the global anomalies as that of a
fundamental spinor representation. Our selection
rule for the generation numbers in is realistically general.

{\it Remark}. For superstring theory with gauge group
$E_8\times E_8'$, a compactification of the heterotic string
is $E_6\times E_8'$, N=1 supersymmetric Yang-Mills theory in
four dimensions$^{42}$. In the $E_6$ sector, the left-handed
Weyl fermions are in the real representation
$\{78\}\oplus 3\times\{27\}\oplus 3\times\{\bar{27}\})\oplus
8\times\{1\}$ of the $E_6$ in terms of the dimensions.
In four dimensions, the $E_6$ is a local anomaly-free group
with $\Pi_4(E_6)={0}$ being trivial. We can show that this
representation will not have a global anomaly when
restricting to a subgroup with non-trivial forth homotopy
group. In this case of $E_6$ upon the reduction to SO(10),
it can also be seen more obviously by our analysis. Upon the
reduction $E_6\downarrow SO(10)$, $78\rightarrow 45\oplus
16\oplus\bar{16}\oplus 1$,
and $27\rightarrow 16\oplus 10\oplus 1$ and correspondingly
for the $\bar{27}$. After the decomposition, there is a
$\bar{16}$ for each 16 and the 10 also appear in pairs.
Therefore, there will be no global gauge anomalies upon
reduction to SO(10). One can also see explicitly that
the theory have no global gauge anomaly for the other
possible gauge subgroups with non-trivial forth homotopy
group. Therefore, the relevant heterotic string theory is
free of both local and global gauge anomalies. An $E_6$
unification gauge theory with even total number for 27 and
its complex conjugate $\bar{27}$ is free of global gauge
anomaly.

{\it Remark}. For the SU(5) and supersymmetric SU(5)
theories, we have shown that the relevant Weyl fermion
representations (e.g. $5\oplus{\bar 10}$) free of local
gauge anomaly are also free of global gauge anomaly for the
gauge subgroups (such as SP(4) and $SU(2)\times SU(2)$
etc.),as well as$^{30, 16}$ SU(2) with non-trivial forth
homotopy group. Moreover, for the relevant representations
of $E_8$ (adjoint or fundamental rep),
SU(4) (for example $8\{4\}\oplus\bar{10}$), and
SU(6) gauge groups free of local gauge anomaly, with
branching rules and the propositions in section 3 it can be
verified that they are free of global gauge anomaly for the
other subgroups with non-trivial relevant homotopy group.
However, we will not present the details further here.
\section{Conclusions}
We have studied the possibilities of global
(non-perturbative) gauge anomalies in a class of gauge
groups. In particular, we investigated the possible
global gauge anomalies in a class of semisimple groups
as products of SU(2) and more generally SP(2N) (N=rank)
groups with non-trivial forth homotopy groups. The results
are applied to the determination of possible global
gauge anomalies in unification groups. Based on the
fact that$^1$ if a gauge theory with Weyl fermions has
global gauge anomalies, then the anomalous or ill-defined
large gauge transformations cannot be unwrapped in quantum
theory when embedded into a larger group G with
$\Pi_4(G)=\{0\}$ although the unwrapping is topologically
well-defined classically, we discussed extensively about
the global gauge anomalies in SO(10) unification theories
containing Weyl fermions in a (f.s) rep. when restricted to a subgroup
$SU(2)\otimes SU(2)\otimes SU(2)\otimes SU(2)$. The physical
consequence is our selection rule$1$
$N_f+N_{mf}=even\ge 4$ for generation numbers in SO(10) and
supersymmetric SO(10) unification theories with $N_f$ and
$N_{mf}$ denoting the family numbers for the ordinary
fermions and mirror fermions. We have also briefly discussed
about the other gauge groups relevant to unification theory
in this connection. Another result is that an odd number of (2,2)
Weyl fermions in $SU(2)\times SU(2)$ (e.g left-right symmetric models)
is inconsistent due to a global gauge anomaly.
We expect that our propositions and the
discussions may be useful for the other general study of
non-abelian gauge theories also.

Finally, we note that the non-trivial forth homotopy group
of SP(2N) gauge groups may induce spontaneous T, or CP,
violations etc. in such gauge theories as noted by the
present author$^{47-49}$. Therefore, our study in connection
to Lie groups is significant to the understanding of gauge
symmetries as well as spacetime symmetries$^{2,50}$.

\section{Acknowledgement}
The author would like to thank S. Okubo, T. Han, Z. Huang,
P. Q. Hung, Y. Li, H. Murayama, and D. D. Wu for valuable
discussions. The author also like to thank W. S. Wilson for
information and discussion about some relevant fifth
homotopy groups as wells as being very helpful for confirming some point of
the topological proof. The author also like to thank
I. Hinchliffe and the theoretical Physics group at LBL for
hospitality when the major work of the paper was done.

\end{document}